\documentclass[prl,twocolumn,superscriptaddress,floatfix]{revtex4-1}
\usepackage{amsmath,amsfonts,amssymb,amsthm,graphics,graphicx,epsfig,bbm}
\usepackage[colorlinks=true,citecolor=blue,linkcolor=blue,urlcolor=blue]{hyperref}
\usepackage[usenames]{color}
\usepackage{graphicx}
\usepackage{subfigure}
\usepackage{epsfig}
\usepackage{bbold}
\usepackage{dcolumn}
\usepackage{bbm}
\usepackage{color}
\usepackage{verbatim}
\usepackage{epstopdf}
\usepackage{amssymb} 
\usepackage{amstext}
\usepackage{latexsym}
\usepackage{hyperref}
\usepackage{amsfonts}
\usepackage{psfrag}
\usepackage{xcolor}
\usepackage{times}
\usepackage{mathtools}

\begin{document}
\title{Genetic Algorithms for Digital Quantum Simulations}
\author{U. Las Heras}
\affiliation{Department of Physical Chemistry, University of the Basque Country UPV/EHU, Apartado 644, E-48080 Bilbao, Spain}
\author{U. Alvarez-Rodriguez}
\affiliation{Department of Physical Chemistry, University of the Basque Country UPV/EHU, Apartado 644, E-48080 Bilbao, Spain}
\author{E. Solano}
\affiliation{Department of Physical Chemistry, University of the Basque Country UPV/EHU, Apartado 644, E-48080 Bilbao, Spain}
\affiliation{IKERBASQUE, Basque Foundation for Science, Maria Diaz de Haro 3, 48011 Bilbao, Spain}
\author{M. Sanz}
\affiliation{Department of Physical Chemistry, University of the Basque Country UPV/EHU, Apartado 644, E-48080 Bilbao, Spain}

\begin{abstract} 

We propose genetic algorithms, which are robust optimization techniques inspired by natural selection, to enhance the versatility of digital quantum simulations. In this sense, we show that genetic algorithms can be employed to increase the fidelity and optimize the resource requirements of digital quantum simulation protocols, while adapting naturally to the experimental constraints. Furthermore, this method allows us to reduce not only digital errors, but also experimental errors in quantum gates. Indeed, by adding ancillary qubits, we design a modular gate made out of imperfect gates, whose fidelity is larger than the fidelity of any of the constituent gates. Finally, we prove that the proposed modular gates are resilient against different gate errors.

\end{abstract}

\date{\today}

\maketitle

\maketitle
Optimization problems, a prominent area in computer science and machine learning~\cite{Alpaydin}, are focused on finding, among all feasible solutions, the best one in terms of efficiency and resource requirements. In particular, genetic algorithms~(GAs)~\cite{Chambers98}, an especially flexible and robust set of optimization methods, are inspired by ideas of evolution and natural selection. In this sense, GAs optimize among different possibilities, which are codified in the genetic information of an individual. Evolution is therefore based on genetic recombination over a group of individuals, together with some random mutations. Natural selection is performed according to the optimization criteria, codified in an evaluation or fitting function. This process is repeated until the individuals satisfy a condition of adaptation. As the solutions to the problem are encoded in the genetic information of the individuals, the information of the survival corresponds to the optimal solution.

A variety of applications have been designed utilizing these methods: mirrors that funnel sunlight into a solar collector~\cite{GrossTED}, antennas measuring the magnetosphere of Earth from satellites~\cite{Hornby06}, walking methods for computer figures~\cite{Geijtenbeek13} and efficient electrical circuit topology~\cite{Zebulum02,Aly10}. The resilience against changes in the initial conditions of the problem is based on the overheads in the resources. For instance, in the case of electric circuits, when one circuit element fails, the circuit continues working and the designed antennas continue measuring signals even under changes in environmental conditions.

One of the most important limitations in the field of quantum computing~\cite{Deutsch85} is the fidelity loss of quantum operations. Quantum error correction protocols~\cite{Shor95, Steane96}, which codify logical qubits in several physical qubits, have been proposed and implemented in different quantum technologies, such as linear optics~\cite{Braunstein98}, trapped ions~\cite{Chiaverini04} and superconducting circuits~\cite{Barends14, Riste15}. It is noteworthy to mention that quantum error correction has been proposed for gate-based quantum computing~\cite{Feynman82} and, in principle, they are also meant to be adaptable to digital quantum simulations~\cite{Lloyd96}. However, experimental implementations of quantum error correction protocols applied to specific quantum algorithms are still to come in the expected development of quantum technologies.

In this Letter, we propose a protocol based on genetic algorithms for the suppression of errors ocurring within digital quantum simulations, along the general lines of bioinspired algorithms in quantum biomimetics~\cite{bioclon,qalife}. First, we prove that GAs are able to decompose any given unitary operation in a discrete sequence of gates inherently associated to the experimental setup. Moreover, we numerically demonstrate that this sequence achieves higher fidelities than previous digital protocols based on Trotter-Suzuki methods~\cite{Lloyd96,Suzuki90}. Second, we show that GAs can be used to correct experimental errors of quantum gates. Indeed, architectures combining a sequence of imperfect quantum gates with ancillary qubits generate a modular gate with higher fidelity than any of the components of the sequence. We exemplify this with a possible implementation of a high-fidelity controlled-NOT (CNOT) modular gate, which is made out of several imperfect CNOTs. Additionally, these architectures show resilience against changes in the gate error. Therefore, by combining the concept of digital quantum simulation with GA, it is possible to design robust and versatile digital quantum protocols.

\paragraph{Genetic algorithms for digital quantum simulations.}

\begin{figure*}[] 
\begin{center}
\includegraphics[width=\textwidth]{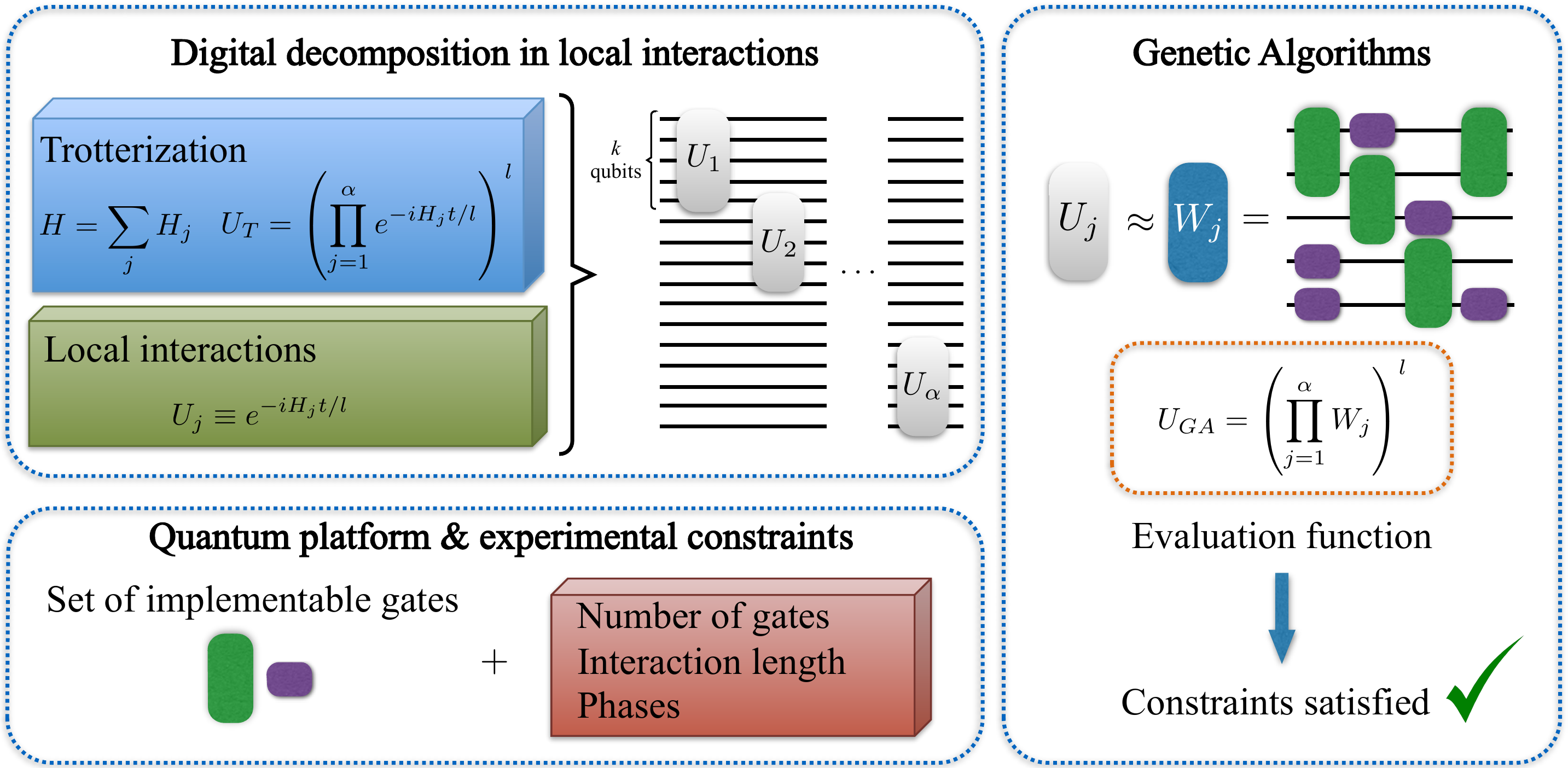}
\end{center}
\caption{Scheme of the GA-based protocol for digital quantum simulations. First, the simulated Hamiltonian is decomposed in local interaction blocks, separately implemented in different unitary evolutions $U_j$ which act on a subset of $k$ particles of the system. Second, the set of gates is selected according to the constraints of the simulating quantum technology: total number of gates to avoid experimental gate error, interactions restricted to adjacent physical qubits, and implementable phases of the Hamiltonian, among others. Once the set of gates is determined, GAs provide a  constraint-fulfilling sequence of gates, which effectively perform the resulting dynamics $U_{GA}$ similar to $U_T$.}
\label{GAProtocol}
\end{figure*}

In the following, we explain how GAs can improve the fidelity of digital quantum simulations. Up to now, the standard technique for realizing digital simulations is Trotter-Suzuki expansion~\cite{Suzuki90}, which has been proven to be efficient~\cite{Berry07,Wiebe11,Sweke16}. This method consists in executing a series of discretized interactions, resulting in an effective dynamics similar to the ideal dynamics of the simulated system. Associated to the unitary evolution of Hamiltonian $H=\sum_j^s H_j$, Trotter formula reads
\begin{eqnarray}
U_I=e^{-iHt} = \lim_{l\to\infty} \left(e^{-iH_1t/l} \cdots e^{-iH_st/l}\right)^l,
\end{eqnarray}
where $U_I$ is the ideal unitary evolution, $t$ is the simulated time, $l$ is the number of Trotter steps, and $H_i$ are the Hamiltonians in the simulating system. On one hand, for a fixed total execution time, the larger the number of Trotter steps is, the lower the digital error of the simulation. On the other hand, the execution of multiple gates in a quantum system can introduce experimental errors due to decoherence and imperfect gate implementation. Therefore, there is a compromise between the number of Trotter steps and quantum operations that can be performed by the quantum simulator~\cite{LasHeras14,Knee15}.

GAs can be employed for outperforming current techniques of digital quantum simulations. The first step of a digital quantum simulation is the decomposition of the simulated Hamiltonian into interactions implementable in the quantum platform, which is a tough task in general. However, by using GAs, it is possible to find a series of gates adapted to the constraints imposed by the quantum simulator, whose resulting interaction is similar to the one of Hamiltonian $H$. For this purpose, we need neither to satisfy the condition $H=\sum_j^s H_j$, nor to use the same execution time for every involved gate. This not only relaxes the conditions for simulating the dynamics, but also allows us to control the number of gates involved, permitting the possibility of minimizing the experimental error.

Let us assume the situation in which is not possible to compute the ideal dynamics of a short-range interacting Hamiltonian, since, for instance the number of particles is too large. By using the Trotter-Suzuki formula, it is possible to decompose the interaction into $\alpha$ local blocks of $k$-interacting particles each, out of $N$ total particles. Let us denote by $U_j$ the ideal unitary evolution of the Hamiltonian acting on the $j$th local block of $k$ qubits. Once the total dynamics is decomposed into blocks, each $U_j$ has to be implemented employing the resources available in the experimental platform, as depicted in Fig.~\ref{GAProtocol}. Here, GAs play an important role, since they provide an architecture for efficiently approximating each $U_j$ by $W_j$:
\begin{eqnarray}
U_{T}&=&\left(\prod_{j=1}^\alpha U_j\right)^l=\left(\prod_{j=1}^\alpha e^{-iH_jt/l}\right)^l,
\end{eqnarray}
\begin{eqnarray}
U_{GA}&=& \left(\prod_{j=1}^\alpha W_j\right)^l,\label{WW}
\end{eqnarray}
where $\alpha=\left\lceil\frac{N-1}{k-1}\right\rceil$.  We assume that $k$ is sufficiently small to allow the minimization of the error associated with the approximation in a standard computer. Therefore, the evaluation function has access to an approximate version of the complete system dynamics, because this is solvable in terms of the Trotter expansion. In our algorithm, as an evaluation function, we compare Trotter unitary evolution, $U_{T}$, for a given number of Trotter steps $l$ with the unitary evolution obtained from GAs, $U_{GA}$. The evaluation function is then given by $R_{j}=||U_{j}-W_{j}||$~\cite{Note1}. In addition, for all analyzed examples, the number of gates involved in the GA protocol is lower than in the Trotter expansion, which gives positive perspectives for experimental realizations of digital quantum simulations based on this approach.

The upper bound for the total error $\xi$ of the protocol, is obtained by combining the Trotter error with the error of the GA optimization $\xi=||U_{I}-U_{GA}|| \le ||U_{I}-U_{T}||+||U_{T}-U_{GA}||$. The first term is nothing but the digital error~\cite{Suzuki90}, so we analyze the second term. Consider that $W_j$, the unitary provided by the GA, has a matrix error $\eta_j$, $W_j=U_j+\eta_j$. Let us denote by $\tilde{U_j}=\mathbb{1}^{\otimes j-1}\otimes U_j \otimes \mathbb{1}^{\otimes\alpha-j}$, the operations when extending to the whole Hilbert space, where $\alpha$ is the number of blocks. The same relation holds for $\tilde{W_j}$ and $\tilde{\eta_j}$, therefore, $\tilde{W_j}=\tilde{U_j}+\tilde{\eta_j}$. We are now able to compute the error of the GA optimization for a single Trotter step, given by $||U_T - U_{GA}||=||\prod \tilde{W_j} - \prod \tilde{U_j}||=||\prod (\tilde{U_j}+\tilde{\eta_{j}}) - \prod \tilde{U_j}||$. We approximate this expression to a first order in $\tilde{\eta}_j$, $|| \sum \tilde{W}_1...\tilde{W}_{j-1}\tilde{\eta}_j\tilde{W}_{j+1}...\tilde{W}_\alpha ||$ $\le$ $\sum ||\tilde{W}_1||...||\tilde{W}_{j-1}||||\tilde{\eta_j}||||\tilde{W}_{j+1}||...||\tilde{W}_\alpha||$. By computing the norm of the unitary matrices $\tilde{W}_j$, we obtain $\sum ||\tilde{\eta_j}||$, which coincides with the error in each of the subspaces, $||U_T - U_{GA}||=\sum||\eta_j||$. Therefore, the GA error is bounded by the sum of the errors in each unitary block, which is linear in the number of qubits for the simulation of a short-range interacting Hamiltonian. As a final remark, since both $W$ and $U$ are unitaries, we would like to point out that the error could also be parametrized by a multiplicative unitary matrix. However, both approaches are equivalent for small errors in the sense that $V_\mu = \exp (i \mu H) \approx 1 + i \mu H + O(\mu^2 ||H||^2)$ for a small $\mu$, so $W \approx U + i U H \mu = U + \eta$.

\begin{figure}[t!]
\includegraphics[width=0.47\textwidth]{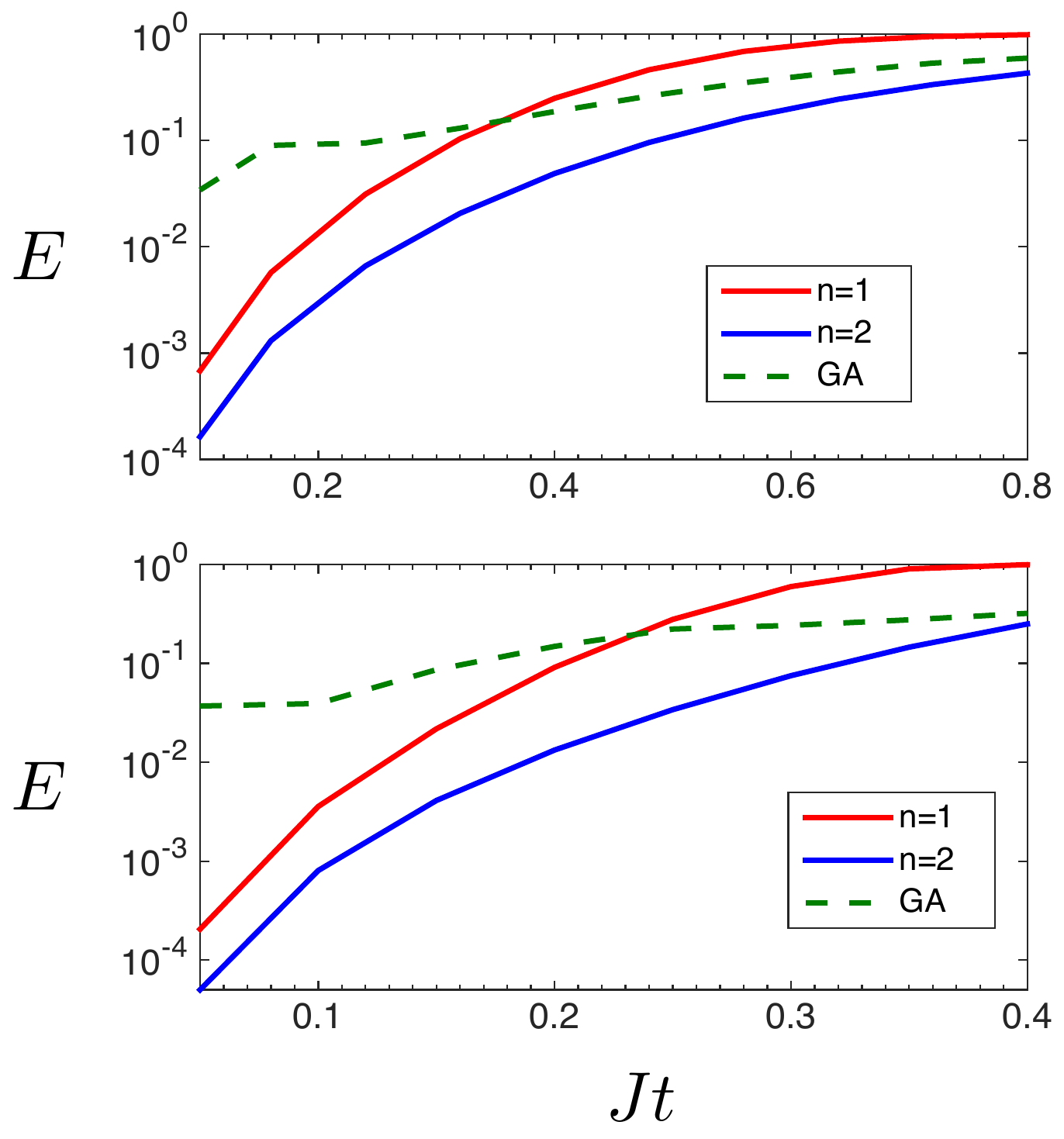}
\caption{Logarithmic plot of the error $E=1-|\langle \Psi|U_I^\dagger \tilde U|\Psi\rangle|^2$ in the evolution of (a) Ising and (b) Heisenberg spin models for $N=5$ qubits, $J=2$, $B=1$, and $|\Psi\rangle=|0\rangle^{\otimes 5}$. Here, $U_I$ is the ideal unitary evolution, while $\tilde U$ refers to the unitary evolution using either a digital expansion in 1 (blue line) and 2 (red line) Trotter steps, or GA (dashed green). The GA protocol requires fewer gates than the digital method for a single Trotter step achieving similar fidelities to two Trotter steps.}
\label{GATrotter}
\end{figure}

We now illustrate the protocol for simulating digitally the isotropic Ising and Heisenberg spin models with a magnetic field in a superconducting circuit architecture~\cite{LasHeras14, LasHeras15, Salathe15}. The Hamiltonians of these models are 
\begin{eqnarray}
H_I&=&J\sum_{\langle i,j\rangle}^N \sigma_i^z\sigma_j^z+B\sum_i^N \sigma_i^x,\nonumber\\
H_H&=&J\sum_{\langle i,j\rangle}^N (\sigma_i^x\sigma_j^x+\sigma_i^y\sigma_j^y+\sigma_i^z\sigma_j^z)+B\sum_i^N \sigma_i^x,
\end{eqnarray}
where $J$ is the coupling between nearest-neighbor spins $\langle i,j\rangle$, $B$ is the strength of the magnetic field, and $\sigma^\gamma_i$ are the Pauli operators acting on the $i$th spin with $\gamma=x,y,z$. We decompose the interactions in terms of single-qubit rotations and controlled-PHASE (CPHASE) gates between nearest-neighbor superconducting qubits~\cite{Barends15a,Barends15b,GarciaAlvarez15}. Following the approach of Ref.~\cite{LasHeras15}, simulating the Ising Hamiltonian requires $N-1$ CPHASE and $3N-2$ single-qubit gates, while Heisenberg Hamiltonian demands 3($N$-1) CPHASE and $11 N-6$ single-qubit gates. In this simulation, we consider a chain of $N=5$ spins. The GA computes a digitalized unitary evolution for a concrete time $t$, constituted by the previous gates in a local subspace of $k=2$ qubits. Then, this unitary evolution $W_1$ is repeated following Eq.~\ref{WW} with $l$=1 over all adjacent qubits due to the translation invariance . The resulting unitary process $U_{GA}$ is compared with the ideal dynamics of the model. This protocol employs $4$ CPHASE and $8$ single-qubit gates for the Ising model, and $4$ CPHASE and $16$ single-qubit gates for the Heisenberg model. Moreover, fidelities are enhanced when compared with the corresponding to pure digital methods for a single Trotter step, even using less gates, as shown in Fig.~\ref{GATrotter}. This approach can be applied similarly to other quantum technologies such as NV centers, trapped ions, and quantum dots among others, just by adding the constrains of their implementable quantum gates to the genetic algorithm. In this protocol, we have considered gates with perfect fidelity. Let us now focus on how to employ GA to improve the experimental error of quantum gates.

\paragraph{Experimental error in a CNOT gate.}

Besides outperforming protocols for digital quantum simulations, GAs are also useful for suppressing experimental errors in gates. We propose a protocol to perform an effective quantum gate by using ancillary qubits and a set of imperfect gates, and we illustrate for the CNOT. A CNOT gate is given by a unitary $U_{CNOT} = \exp (i \frac{\pi}{2} H_{CNOT})$, with $H_{CNOT} = \frac{1}{2}\left[(\mathbbm{1} + \sigma^z)\otimes\mathbbm{1}+(\mathbbm{1}-\sigma^z)\otimes\sigma^x\right]$. Let us consider imperfect gates modeled by $W_{CNOT} =  \exp (i \frac{\pi}{2} H_{CNOT} + \delta H_R)$, with $\delta << 1$ and $H_R$ a random matrix, such that $||H||_2 = 1$. These unitaries define unital quantum channels $\mathbbm{E}_{U}= U \otimes \bar{U}$ and $\mathbbm{E}_{W}= W \otimes \bar{W}$ respectively, and we define the error of the gate as the distance $\eta=||\mathbbm{E}_{W}-\mathbbm{E}_{U}||_2$. 

Let us now consider $q-2$ ancillary qubits in the state $|0\rangle$ in addition to the control and target of the integrated CNOT. Let us also consider $n$ imperfect CNOT gates $\vec{W} = \{ W_1 , \ldots , W_n \}$ acting on any possible pair of the $q$ qubits, with errors $\vec{\eta} = \{\eta_1, \ldots , \eta_n \}$ respectively, and denoted by $\eta = \min_i \eta_i$. The integrated circuit is defined by a set of $n$ ordered pairs $IG_{\vec{\eta}}=\{ (i_k,j_k) | 1 \leq i_k, j_k \leq q , k=1,\ldots,n \}$, where the indices indicate the control and target qubits, respectively. In order to calculate the fidelity of the protocol, we compute the Kraus operators of the integrated CNOT, by tracing out the $q-2$ ancillary qubits, and compare the resulting channel $\mathbbm{E}_{IG_{\vec{\eta}}}$ with the unital channel $\mathbbm{E}_{U}$, $\epsilon_{IG_{\vec{\eta}}} = ||\mathbbm{E}_{IG_{\vec{\eta}}}-\mathbbm{E}_{U}||_2$. If $\epsilon_{IG_{\vec{\eta}}} < \eta$, then the CNOT gate is implemented with higher fidelity than any of the original CNOT gates, showing this GA-based architecture resilience against quantum errors. 

The set $IG_{\vec{\eta}}$ codifies the genetic information of the individuals which conform the population evolving into successive generations. During the reproduction, the individuals recombine their genetic code, which is also allowed to mutate. The survival probability depends on the fidelity of the effective CNOT encoded in $IG_{\vec{\eta}}$ and, therefore, only individuals associated with a small error succeed.

The number of possible architectures involving $n$ different CNOT gates and $q$ ancillary qubits is $P=(q^2-q)^n n!$~\cite{supp}. The factor $(q^2-q)^n$ is due to all possible CNOT configurations in a given order between qubits $i$ and $j$ for $n$ gates, while $n!$ comes from reordering imperfect gates $\{W_1,...,W_n\}$. When $q$ and $n$ are small, the optimal architecture can be found by analyzing all cases. However, when we increase these parameters, this brute-force optimization method turns out to be inefficient. GAs allow us to optimize the protocol in this unreachable regime, being moreover robust, as analyzed below.

\begin{figure}[h!]
\includegraphics[width=0.47\textwidth]{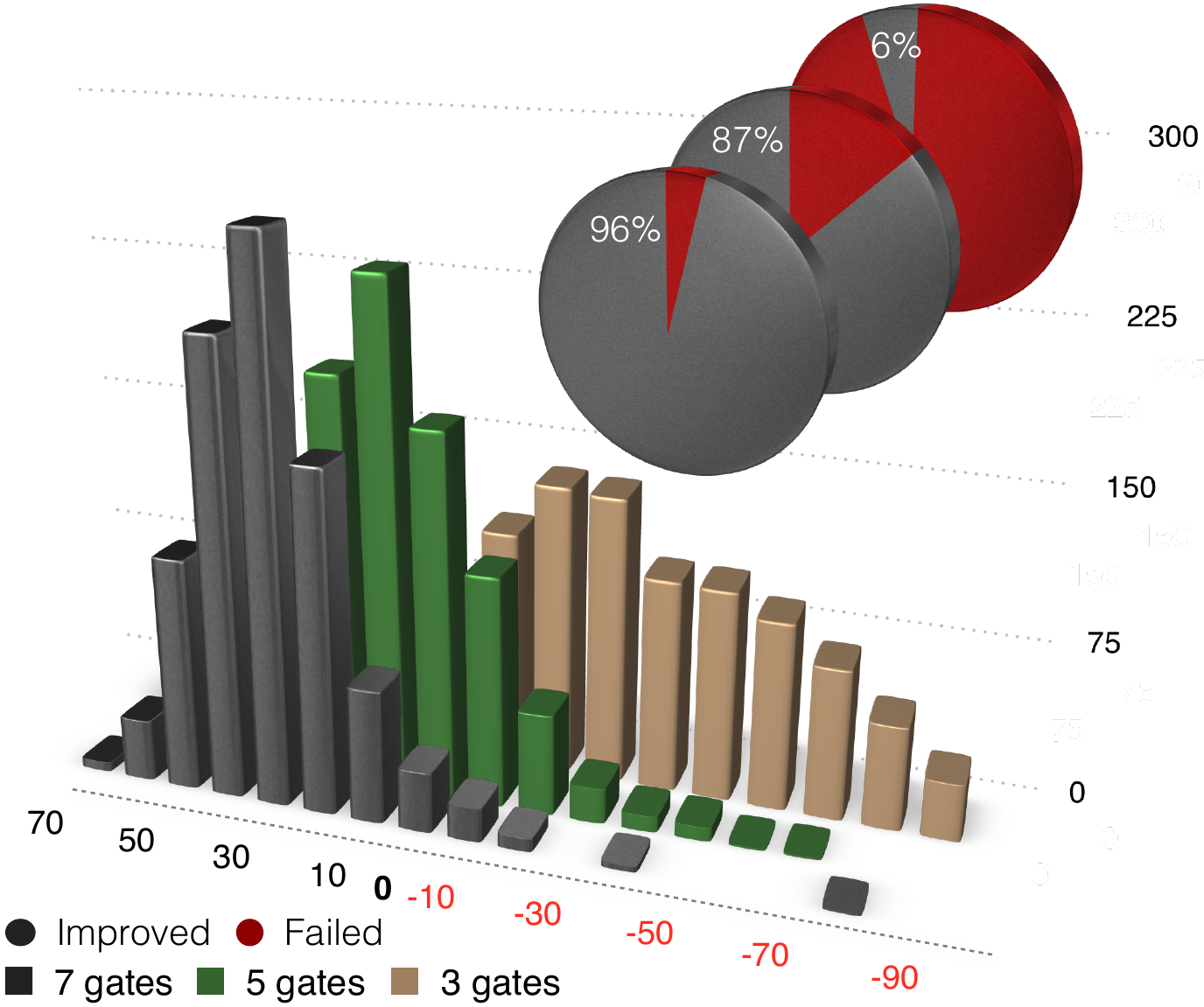}
\caption{Error resilience for architectures with $n=3,5,7$ imperfect CNOT gates using 1000 runs. Pie charts show the percentage of cases in which the fidelity of the effective CNOT overmatches the best CNOT employed in the architecture.
Bar charts show the distribution of cases according to the relative improvement in the error, again when compared with the best CNOT.} 
\label{Bars}
\end{figure}

\begin{figure}[h!] 
\centering
\includegraphics[width=0.30\textwidth]{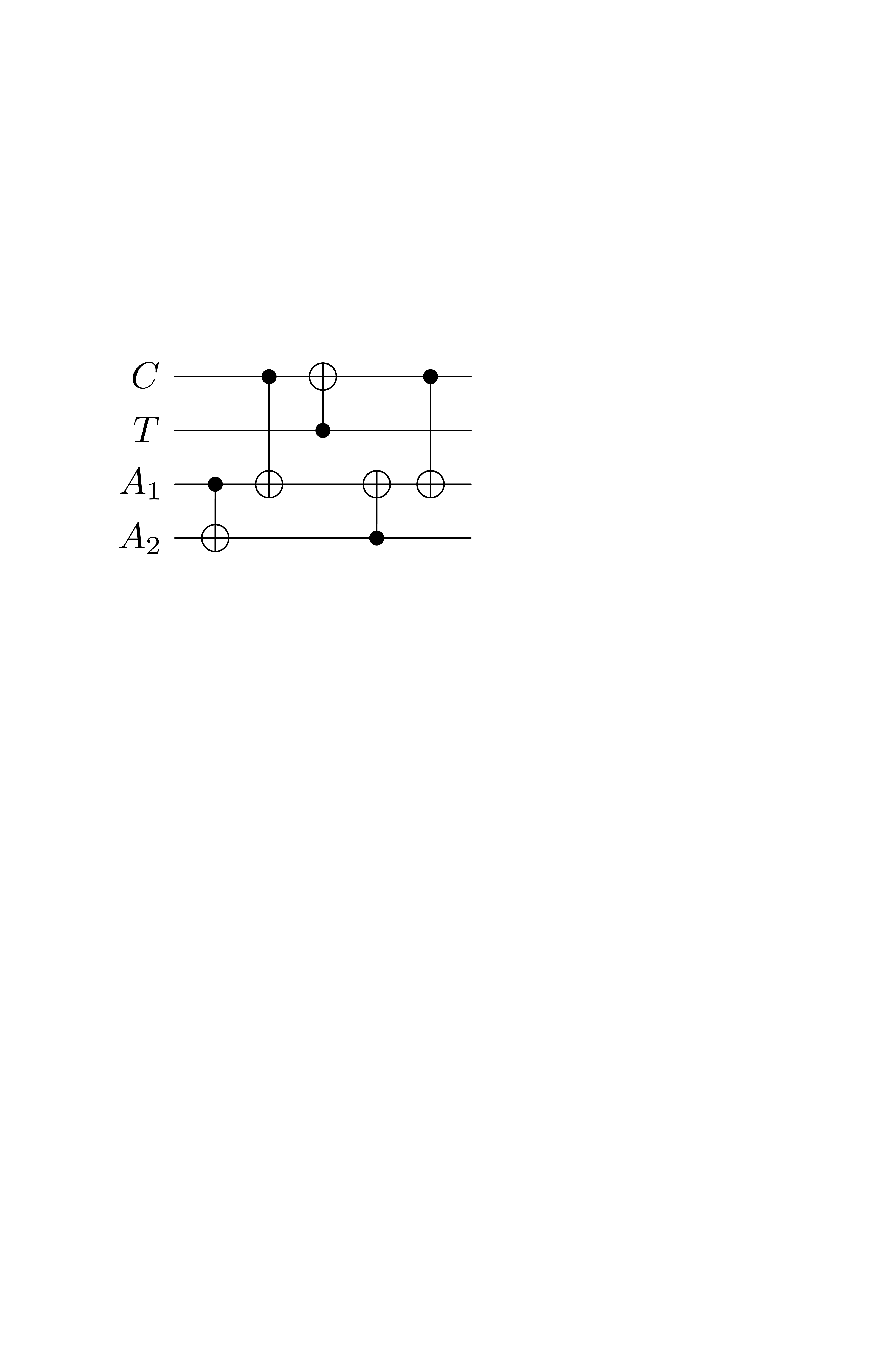}
\caption{Scheme of the optimal architecture for constructing a CNOT gate with $5$ imperfect gates, by using two ancillary qubits initialized in state $|0\rangle$. Here, $C$ is the control, $T$ is the target, and $A_1$ and $A_2$ are the ancillary qubits. } 
\label{fig1}
\end{figure}

This CNOT case has been analyzed involving three, five and seven gates. Notice that, when one considers $q=4$ and $n=7$, the number of possible architectures is larger than $1.8\times, 10^{14}$ for a fixed set of imperfect gates. We have chosen a set of gates and find the optimal architecture by GA. Then, we analyze the resilience or robustness of this architecture by changing the set. In Fig.~\ref{Bars}, we have depicted the results for a sampling of $1000$ sets of random imperfect CNOT gates. The pie charts show the percentage of cases with a lower error than any CNOT performed in the protocol, which are $6\%$ for three qubits, $87\%$ for five, and $96\%$ for seven. Furthermore, the bar charts show the average improvement of the error for the integrated CNOT with respect to the best implementing CNOT, which is $-39\%$, $+18\%$, and $+30\%$, respectively. For completeness, in Fig.~\ref{fig1}, we show the optimal architecture for $q=4$ and $n=5$, obtained from a fixed set of imperfect gates $\vec{W}$, and proven to be robust \cite{supp}.

Additionally, we have studied the behavior of the protocol with respect to the number of ancillary qubits. The results show no significant improvement when the number of performed gates is small \cite{supp}. For instance, architectures up to $n=7$ do not overcome fidelities shown above when adding a third ancillary qubit, $q=5$. However, we expect that architectures with a larger number of gates would actually take advantage of using more ancillary qubits in order to suppress the error.

The same protocol can be applied in the realization of more general unitary operations. Additionally, the gates conforming the building blocks can be arbitrary, which facilitates the adaptation of the protocol to any experimental platform.

In summary, in this work we proposed a new paradigm based on GAs to enhance digital quantum simulations and face different types of quantum errors. We showed that they can be used to improve the fidelity of quantum information protocols by effectively reducing digital errors produced in Trotter-Suzuki expansions. Our method allowed us to correct experimental errors due to imperfect quantum gates, by using ancillary qubits and optimized architectures. We also argued that solutions provided by GAs manifest resilience against digital and experimental quantum errors. From a wide perspective, we expect that GAs will be part of the standard toolbox of quantum technologies, and a complementary approach to analog \cite{Nebendahl09, Choi14} and digital \cite{Machnes11} optimal-control techniques.

We acknowledge useful discussions with Rami Barends and Ryan Sweke, and support from Spanish MINECO FIS2012-36673-C03-02 and FIS2015-69983-P; UPV/EHU UFI 11/55 and a Ph. D. grant, Basque Government IT472-10 and BFI-2012-322; and SCALEQIT EU projects. E.~S. also acknowledges support from a TUM August-Wilhelm Scheer Visiting Professorship and the hospitality of Walther-Meissner-Institut and TUM Institute for Advanced Study.

\section{SUPPLEMENTAL MATERIAL}\label{sec:chip}
In this Supplemental Material, we discuss details useful for the understanding of the main results of the paper.

\subsection*{Description of the Genetic Algorithm}
In this section, we describe the GA \cite{Alpaydin, Chambers98} used to obtain the decomposition of the local Trotter blocks \cite{Lloyd96,Suzuki90}. The sequence of quantum gates is codified in a matrix representing in the protocol the genetic code of an individual. This matrix contains as many columns (genes) as allowed resources, and sufficient rows to determine the type of gate and the qubits on which it acts. The next step is to engineer a fitness or evaluation function which maps every individual into a real number. This allows to classify the individuals with respect to an adequate criterion for the optimization purposes. In our case, the fitness function corresponds to the fidelity with respect to the ideal block dynamics. Finally, each cycle of the algorithm consists of three stages: breeding, mutation, and natural selection. 

In the breeding stage, a new generation of individuals is obtained by combining the genetic code of the predecessors, which provides the genetic code of the offspring. We have used a hierarchical combination method, which allows the number of broods of each individual to depend on its fidelity. In particular, for an initial population of $4$ individuals sorted by fidelity, our algorithm creates an offspring of $9$, $6$ of which acquire genetic material of the first precursor, $5$ of the second, $4$ of the third, and $3$ of the fourth. Notice that each newborn individual is produced with the genetic information of two predecessors, as it can be seen from the fact that adding the numbers of each progeny equals two times the number of newborn individuals. Notice that this is not the most general situation, since we could have considered individuals as a combination of more predecessors. Additionally, the amount of genes each precursor provides, in this case the number of matrix columns, also depends on the hierarchy induced by fidelity. 

In the mutation stage, every individual is allowed to mutate by randomly modifying any sequence of genetic material, with equal probability for all individuals. This probability settles the threshold to overcome for a random number for a mutation event to occur, case in which another set of random numbers provides the new genes to insert in the genetic material. 

In the last stage of the cycle, old and new generations of individuals are combined in the same population group. Afterwards, they are sorted depending on their fidelity, and those which show the highest fidelity are selected as the initial population of the forthcoming cycle. 

We have observed that it is convenient to combine numerical trials with high and low mutation rates to enhance the breeding or the mutation stages depending on the intermediate results.  

\subsection*{Number of Architectures}
We derive here the formula $P=(q^2 -q)^n n !$ for the number of architectures in terms of the number of ancillary qubits $q$ and the number of imperfect gates $n$. We impose the condition of applying each two-qubit gate once and only once, and that the gates are asymmetric, so applying it to qubits $(i,j)$ is different to apply it to qubits $(j,i)$. Therefore, one of the $q$ qubits is selected as the control, and one of the remaining $q-1$ as the target. This process is repeated for each of the $n$ gates, so we obtain $(q(q-1))^n$ possibilities. Finally, the $n$ gates may be applied in any possible order, so there are $n!$ re-orderings. Therefore, by combining both results, the number of total architectures turns into $n! \,(q^2-q)^n$.

\subsection*{Errors in architectures building the CNOT gate}
We compare the mean error of the integrated CNOT gate obtained with GA over many realizations of imperfect gates with the average of the highest fidelity imperfect CNOT gate involved in the architecture. For this purpose, we take a sampling of 1000 different experiments, and we average the error of the best gate. We estimate the error of the integrated CNOT and obtain the percentage of improvement in the error. These results are summarized in Table \ref{Table} for the cases studied in the main manuscript. As it is shown, the probability to have a high-fidelity gate is increased when the number of gates is aucmented. Accordingly, there are more possible architectures that minimize the error in the integrated CNOT. For the case of $q=5$ and $n=7$, we obtain similar errors to the ones for $q=4$. This could well be because the number of ancillary qubits is of the same order of the involved gates, and then no measurable improvement is expected since there is no cancellation of gate errors. Nevertheless, the optimal relation between number qubits and involved gates is still an open question.

\begin{widetext}
\begin{center}
\begin{table}[h!] 
\begin{tabular}{{|l|l|l|l|l|}}
\hline
	& $q=4, n=3$ & $q=4, n=5$ & $q=4, n=7$ & $q=5, n=7$\\
\hline
  Error of best gate & 0.1271& 0.1205 & 0.1150 & 0.1150 \\
\hline
  Error of architecture & 0.1771& 0.0988 & 0.0807 & 0.0810 \\
\hline
Approximate improvement & -39$\%$ & 18$\%$ & 30$\%$ & 30$\%$\\
\hline
\end{tabular}
\caption{Average errors of integrated CNOTs and highest fidelity CNOTs for the protocols involving $q$ qubits and $n$ gates.}\label{Table}
\end{table}
\end{center}
\end{widetext}

\end{document}